\documentclass[useAMS]{mn2e}
\usepackage{mnras_cite}
\usepackage{epsfig}
\usepackage{color}

\usepackage{graphicx}
\usepackage{subfigure}

\newcommand{\dif}{{\rm d}}

\begin{document}

\title[The formation of galaxy disks in a hierarchical Universe]{The formation of galaxy disks in a hierarchical universe} 
\author[M.~J.~Stringer, A.~J.~Benson]{M.~J.~Stringer$^{1,2}$, A.~J.~Benson$^1$\\ 
1. Theoretical Astrophysics, Caltech, MC130-33, 1200 E. California
Blvd., Pasadena, CA 91125, U.S.A. (e-mail: abenson,stringer@caltech.edu)\\ 
2. Department of Astrophysics, Keble Rd., Oxford, OX1 3RH, U.K.\\ 
}

\maketitle

\begin{abstract}
The formation of galactic disks and the efficiency of star formation within them are issues central to our understanding of galaxy formation. We have developed a detailed and versatile model of disk formation which combines the strengths of previous studies of isolated disks with those of hierarchical galaxy formation models. Disk structure is inferred from the distribution of angular momentum in hot halo gas and the hierarchical build-up of dark matter, leading to theoretically generated systems where the evolution of surface density, rotation, velocity dispersion, stability and metallicity is predicted for annular regions of width 20--100pc. The model will be used to establish whether the accepted theory of large scale structure formation in the Universe is consistent with observed trends in the properties of disk galaxies.

This first paper explicitly examines the importance of embedding such calculations within a merging hierarchy of dark matter halos, finding that this leads to dramatically different formation histories compared to models in which disks grow in isolation. Different models of star formation are explored, and are found to have only a secondary influence on the properties of the resulting galaxy disks, the main governing factor being the infalling gas supply from the hot halo.
\end{abstract}

\section{Introduction}\label{Introduction}

Stellar disks are one of the most conspicuous features of galaxies: they are the sites of quiescent star formation and display such distinctive features as spiral arms and bars. Current data suggest that 40--60\% of all luminosity in the local Universe is contributed by stars located in galaxy disks \cite{Tasca05,Benson07}. In cold dark matter universes, where structure forms hierarchically, it is believed that galactic disks are the first components to form, with elliptical galaxies and bulges of spiral galaxies forming later through the merging of pre-existing galaxy disks \cite{Kauffmann93,Baugh96}. Understanding in detail how galaxy disks form and evolve is therefore crucial to our understanding of galaxy formation as a whole.

Traditionally, semi-analytic models of galaxy formation have estimated the total masses of stars, gas, and metals contained in each galaxy disk, but not the radial distribution of these components (see \pcite{Kauffmann99,Somerville99,Cole00,Hatton03} for recent examples). Instead, they explicitly assume that the surface density has an exponentially declining radial dependence. Consequently, such models are forced to adopt highly simplified prescriptions for star formation, which are unable to capture the physical mechanisms involved in this complex process. 

Conversely, models of individual disks (e.g. \pcite{MMW98a,MMW98b,Firmani00,Efstathiou00,vdB01,Monaco04,Marios06}) study the radial distribution of these properties in great detail but do not always consider the dynamical evolution of the system, digressions from the steady state or, perhaps more importantly, interactions with neighbouring galaxies. 

A similar dichotomy has existed within simulations of galaxy formation, which have often focused on either the detailed evolution of a single halo \cite{Abadi03} or on the properties of large samples of galaxies using comparatively low resolution \cite{Navarro00}. This situation can be addressed by introducing methods to increase resolution in areas of interest, beginning on cosmological scales ($\sim100$Mpc) but selecting smaller regions ($\sim5$kpc) to study using a softening lengths as small as 500pc \cite{SL03,Governato04,Robertson04}.

In order to make progress, the respective strengths of all these approaches must be brought together. Given sufficient information about the halos in any analytic or numerical model of large-scale structure, the profile of the disk which forms within each sub-halo can be determined with simple assumptions for the symmetry of the system and the conservation (or otherwise) of the angular momentum of halo gas as it cools and condenses to form a disk. Local values of the star formation rate, the rate of outflow of gas, metallicity and the velocity dispersion of gas clouds and so forth can then be estimated as a function of radius in the disk.

This approach immediately provides three distinct advantages over previous work. Firstly, it allows the surface density profiles of galaxy disks to be determined from more fundamental premises (the distribution of halo gas, angular momentum conservation, the frequency of galaxy mergers and so on) rather than simply by assumption. The metallicies of stellar and gaseous components can be followed separately, rather than assuming that they trace the overall density. Secondly, knowledge of the surface density profile and associated rotation curve and stability allows for the inclusion of star formation models based upon local density, local stability on so on. Finally, all of this can be modelled within a realistic hierarchical galaxy formation scenario.

We have developed such an improvement within the {\sc Galform} semi-analytic model \cite{Cole00} and the remainder of this paper describes the techniques employed (\S\ref{Model}) and presents basic results (\S\ref{Applications}).The effects of hierarchical structure formation and galaxy mergers on disk formation are studied using three alternative models for star formation. These comparisons are carried out initially in the simplified context of an isolated, individual dark matter halo and then in the context of a similar system formed within a fully hierarchical model of galaxy formation. Conclusions are drawn in \S\ref{Conclusions}.

\section{The Model}\label{Model}

This section describes the methods used to compute the radial structure and star formation within galaxy disks. The model can be embedded within the {\sc Galform} semi-analytic model of galaxy formation---which is not described in detail here---although it can also be run independently of {\sc Galform}. The reader is referred to \scite{Cole00} and \scite{Baugh05} for complete details of {\sc Galform}. Only those parts of the model directly affected by recent modifications are detailed here. Throughout this work, a standard $\Lambda{\rm CDM}$ model of cosmology is assumed, with values chosen for consistency with the latter publication ($\Omega_{\rm M}=0.3, \Omega_\Lambda=0.7, \Omega_{\rm b}=0.04,\sigma_8=0.93, h_0=0.7$).

\subsection{Halo Structure}\label{Halo}

In the $\Lambda$CDM cosmogony, the mass density of the Universe is dominated by cold dark matter, which interacts only via gravity \cite{DMreview}. Simulations of the behaviour of such particles \cite{NFW97} show that they undergo gravitational collapse into haloes with radial density dependance:
\begin{equation}\label{NFW}
\rho(r) = \frac{\rho_{\rm v}}{\frac{cr}{r_{\rm v}}\left(1+\frac{cr}{r_{\rm v}}\right)^2} .
\end{equation}
The virial radius, $r_{\rm v}$, encloses the correct mean density for spherical top-hat collapse in the particular cosmology, thus specifying $\rho_{\rm v}$. For the cosmology used, this is 330 times the mean density at $z=0$ \cite{Eke96}. The only free parameter, $c$, is found to be strongly correlated (and decreasing with) total halo mass \cite{NFW95,Bullock01}, albeit with a large scatter.

The pressureless dark matter particles are supported against further collapse only by their orbital motion. As baryonic material cools and condenses towards the centre of the halo, the dark matter density profile will be adiabatically compressed. Following \scite{Blumenthal86}, we model this compression by assuming that dark matter particles conserve the adiabatic invariant quantity $j$ during the process of baryonic collapse and that there is no shell crossing ($\dif M(j)/\dif t=0$). Throughout this paper is that $M(j)$ and $M(r)$ always refer to the mass of material with radius less than $r$ or specific angular momentum less than $j$.

This calculation uses $j=\sqrt{GMr}$ for the initial dark matter distribution. Simulations of dark matter contraction on galactic scales \cite{Gnedin04} have indicated that this approximation overestimates the central density by not allowing for orbital eccentricity in the dark matter. More recently however, \scite{Dutton07} made additional corrections for the flatness of the disk and found that the combined effects lead to little net change from the initial, basic assumption. We therefore retain this simple approach.

The baryonic component is assumed to be shock heated to the halo virial temperature during collapse. This hot gas is supported by thermal pressure and so, as it subsequently begins to cool, it can fall towards the halo centre. During this infall we assume that the collective {\it drift} angular momentum of the gas about some common axis is conserved (this angular momentum arises as a result of tidal interactions between halos; \pcite{Hoyle49}). This claim has not always been supported by simulations \cite{Navarro94,Navarro00} though recent work \cite{Governato04} indicates that this may well have been an issue of resolution. The assumption of conservation of angular momentum during collapse is one that can be changed within our model, a flexibility that will be used to advantage in future work.

\begin{figure}
\centering
\includegraphics[trim = 15mm 150mm 15mm 35mm, clip, width=\columnwidth]{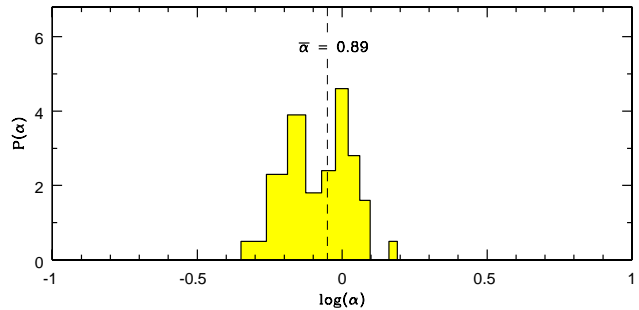}
\caption{Values of the parameter $\alpha$ from a sample of 41 halos in a smoothed particle hydrodynamics (SPH) simulation \protect\cite{Sharma05}. This parameter controls the shape of the angular momentum distribution throughout the halo, as described by eqn.~(\ref{haloAM}).}
\label{AlphaDist}
\end{figure}

The distribution of density and angular momentum for simulated hot gas halos was recently found to closely match the following forms \cite{Sharma05}:
\begin{equation}\label{haloAM}
\rho_{\rm g}(r) \propto \left(1 + \frac{r}{r_{\rm H}}\right)^{-3}~~;~~ \frac{\dif M(j)}{\dif j} \propto j^{\alpha-1}\exp\left(-\frac{\alpha j}{\langle j\rangle}\right).
\end{equation}
Both expressions are normalized so that the appropriate total baryonic mass, $(\Omega_{\rm b}/\Omega_{\rm M})M_{\rm v}$ (where $M_{\rm v}$ is the virial mass of the halo), is contained within the virial radius, as is the total angular momentum:
\begin{equation}
J_{\rm tot} = M_{\rm v}\langle j\rangle = \lambda {\rm G} M^{5/2}/|E|^{1/2}.
\label{J_total}
\end{equation}
This defines the {\it spin parameter}, $\lambda$. The value of this parameter, of $\alpha$ and $r_{\rm H}$ are particular to each halo, being distributed around typical values of 0.03, 0.9 and $r_{\rm v}/20$ respectively. Figure~\ref{AlphaDist} shows the distribution of $\alpha$ found by \scite{Sharma05}. 

To relate the angular momentum distribution to the radially dependent density, we follow \scite{Efstathiou00} by assuming that {\em cylindrical} shells in the halo share the same $j$. Converseley, \scite{Firmani00} and \scite{vdB01} both assume that the two components have the same density distribution at virialisation and that each {\em spherical} shell is in solid body rotation.
\pagebreak
\subsection{Disk Formation}\label{DiskFormation}
Though the details of non-linear gravitational collapse are most accurately modelled by N-body simulation, continuing to track the three dimensional location of the gas particles as they cool is computationally impractical as part of a cosmological-scale model (the resolution achieved by \scite{Knebe06}, for example, is of the order of a kpc -- too large to resolve individual galaxies accurately). By using the fact that the system has non-zero angular momentum and hence a well defined axis, the cooling phase can be modelled assuming cylindrical symmetry. This greatly reduces the complexity of further calculations without excessive departure from a realistic physical picture.

We use the cooling model described in detail by \scite{Cole00}, which uses the cooling function $\Lambda(T) = \dot{\varepsilon}/n^2$ calculated by \scite{Sutherland93}. Here $T$ is the gas temperature, $\dot{\varepsilon}$ is the energy loss rate per unit volume and $n$ is the number density of hydrogen. The time taken for gas particles to cool is therefore
\begin{equation}
t_{\rm cool}(r) = \frac{3}{2}\frac{\mu {\rm m}_{\rm H}}{\rho_{\rm g}(r)}\frac{{\rm k}_{\rm B}T}{\Lambda(T,Z)},
\end{equation}
where gas density is $\rho_{\rm g}$, mean molecular mass is $\mu m_{\rm H}$ and $k_{\rm B}$ is Boltzmann's constant. As it cools, the gas will lose pressure, falling out of hydrostatic equilibrium and moving towards the centre of the halo until its residual angular momentum lends sufficient rotational support against further collapse \cite{FallEfstathiou80}. The time taken for this to happen is assumed to be the greater of $t_{\rm cool}$ and the freefall time,
\begin{equation}
t_{\rm ff}(r) = \int_0^r \left[ \int_r^{r^{\prime\prime}} - {2{\rm G}M(r^\prime)\over r^{\prime 2}}{\rm d}r^\prime\right]^{-1/2}{\rm d}r^{\prime\prime}.
\end{equation}
In the idealized situation that the angular momenta of all the gas shells are mutually aligned, the result will be a thin disk of gas with the original angular momentum vector normal to the disk plane.

Both gas and dark matter components are assumed to conserve their specific angular momenta during the collapse process. Therefore, all baryonic material and dark matter which has the same initial specific axial angular momentum will share a {\it common final radius} in the plane which can be estimated by\footnote{An important underlying assumption is that the specific angular momentum of the halo gas increases monotonically with its distance from the axis.}:
\begin{equation}
R(j) = \frac{j^2}{{\rm G}\left[M_{\rm disk}(j) + M_{\rm DM}(j)\right]},\label{R_basic}
\end{equation}
where $j$ is the specific angular momentum of circular orbits at radius $R$ in the disk plane and $M_i(j)$ is the mass of each component with specific angular momentum less than $j$. 

The disk is broken up into 200 concentric annuli between the origin and the final radius of the most recently cooled gas. These are spaced linearly in specific angular momentum. After each timestep, every annular region acquires all the halo gas within two cylindrical bounds, given by the range in specific angular momentum of that region, and two spherical bounds enclosing the portion of the halo which has just cooled. After all masses have been updated, the function $R(j)$ is reevaluated resulting in a revised profiles $M_{\rm disk}(R)$ and $M_{\rm DM}(R)$.
The masses, mean metallicities and velocity dispersions of stars and gas can then be tracked for each of these regions. As such, individual annuli have widths of 20--100pc in typical disks. This choice is made so that regions are large enough for the use of scaling laws and locally averaged values to be appropriate, yet small enough so that a detailed picture of the disk profile is produced. Appendix~\ref{CompDetails} shows that the final results are not particularly sensitive to this choice.

Following thousands of disk profiles in such detail is computationally impractical and, as it turns out,  largely unnecessary. The properties of each disk as a function of specific angular momentum are therefore mapped onto a new grid of only 20 annuli after each calculation of gas infall and star formation. This coarser grid is used for the rest of the calculations (e.g. for computing galaxy properties during mergers, see \S\ref{Mergers}). Increasing this number above 20 produces negligible change in the shape of the eventual profiles, as illustrated in Appendix~\ref{CompDetails}. The full set of 200 annuli is then recreated by interpolation within this coarser grid to calculate gas cooling, star formation and outflow in each particular disk.
\begin{figure*}
\begin{tabular}{cc}
\includegraphics[trim = 8mm 56mm 12mm 25mm, clip, height=1.47\columnwidth]{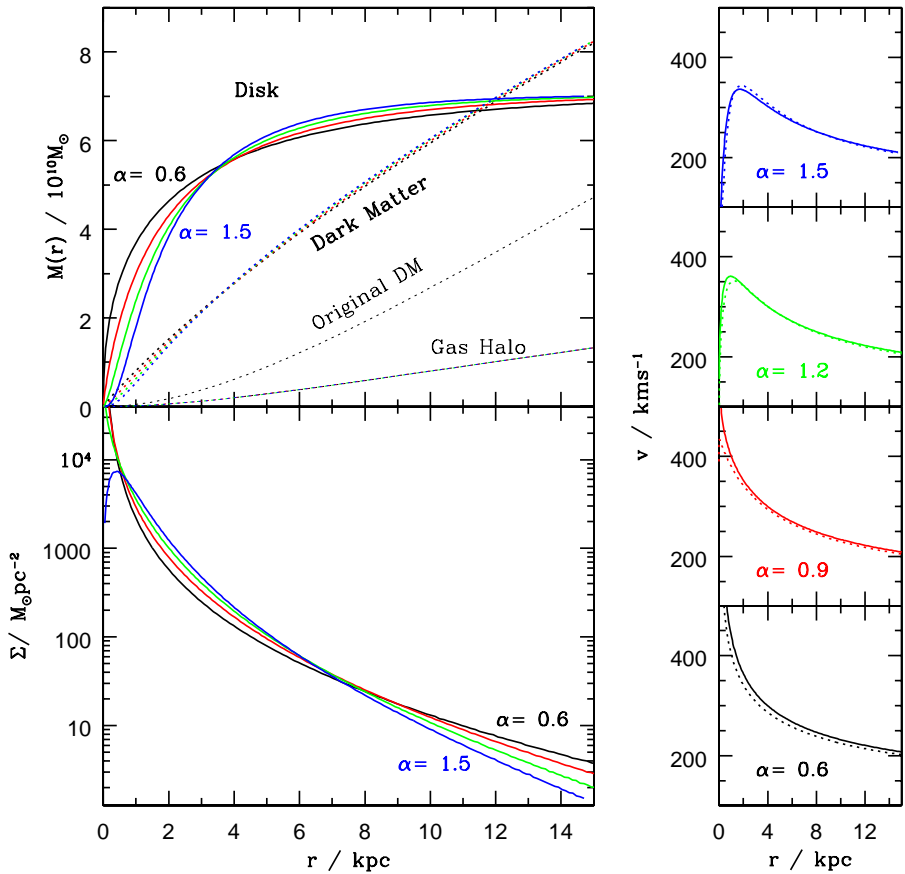}
\end{tabular}
\caption{Galactic disks which would  form if all the gas in a $5\times10^{11}M_\odot$ virialised halo were to cool onto a plane, conserving the net component of its initial angular momentum as it does so. This constitutes a total baryonic mass of $7\times10^{10}M_\odot$. in this example, the spin parameter, $\lambda=0.03$ and dark matter contraction parameter $c=10$. The top left panel shows the mass of all the different components enclosed at each radius with surface density shown in the lower left panel.  Circular velocity is plotted on the right. Original density and angular momentum profiles are given in equations (\ref{NFW}) and (\ref{haloAM}).  Each line corresponds to a slightly different value of $\alpha$ (which governs the distribution of angular momentum in the original halo). The solid lines in the rotation curves plots were calculated using the approximation $j^2=GMR$, while the dotted lines use equation (\ref{DiskIntegral}) which approximates the disk as being perfectly thin.}
\label{alpha}
\end{figure*}

Clearly eqn.~(\ref{R_basic}) incorrectly approximates the potential due to the disk as being spherically symmetric.  The alternative option, treating the disk as perfectly thin, is also only an approximation to reality and considerably complicates the calculations, mainly because there is no analytic form $R(j,M)$. The relation between the three variables in this case would be \cite{Cuddeford93}:
\begin{equation}
j^2(R) = \frac{4G R^2}{2\pi}\int_0^R\frac{a\dif a}{\sqrt{R^2-a^2}}\frac{\dif}{\dif a}\int_a^\infty\frac{\dif M(R')}{\sqrt{R'^2-a^2}}.\label{DiskIntegral}
\end{equation}
An initial estimate of the mass profile $M(R)$ then has to be corrected iteratively until this equation and (\ref{R_basic}) are both satisfied. 

As shown in Fig.~\ref{alpha}, there is surprisingly little difference in results between the two approximations. Though they would differ by as much as 20\% for an exponential disk, the overriding contribution of the dark matter and the relative central concentration of these disks reduce the typical discrepancy to less than 5\%. The less computationally demanding approximation, eqn.~(\ref{R_basic}), will therefore be used throughout this work. 

Figure~\ref{alpha} also shows the corresponding mass and surface density profiles of the disk (along with mass profiles for the original gas halo and dark matter) for each value of $\alpha$.  One clear feature to note at this stage is the sharp rise in surface density near the axis, leading to rotation curves with central peaks far greater than those observed, particularly for lower values of $\alpha$.

This feature is also seen in the model used by \scite{Dalcanton97}, which treats the halo as a uniformly rotating sphere. Given the simplicity of that assumption, the authors claimed no profound importance for these bulge-like centers. In the present case, the initial conditions are derived instead from simulations of halo formation and their lack of low-radius resolution provides a similar source of uncertainty.

This problem of excess low angular momentum material persists even in simulations of disk systems \cite{Abadi03} though, encouragingly, increased resolution does seem to lead to more observationally consistent mass distributions \cite{Governato06}.

\scite{Firmani00} and \scite{vdB02} investigate this issue in detail, finding that systems with a lower baryon mass fraction (0.04-0.09) or higher spin (0.03-0.08) have more observationally consistent rotation curves. The factors responsible for the excessively peaked rotation curves here are therefore (1) a halo spin parameter of 0.03, (2) complete cooling of the baryonic component and (3) a universal baryon mass fraction of 0.13.

Each of these differences are justifiable: Halos with $\lambda>0.03$ are rare in the simulations, so that value is appropriate. \scite{vdB02} finds that the fraction of the cooled baryonic mass approaches one for halo masses below $\sim10^{11}M_\odot h^{-1}$ suggesting that complete cooling may not, in fact, be entirely unphysical for this halo. Lastly, the choice of cosmological parameters here actually {\em underestimates} the baryon fraction by comparison with the latest values from WMAP of $\Omega_{\rm M}=0.29$ and $\Omega_{\rm b}=0.047$ \cite{WMAP}. In conclusion, these choices are sufficiently well motivated to suggest that there may indeed be some missing physics.

One obvious candidate is gas outflow due to supernova winds but, when this is taken into account in \S\ref{Applications}, this model suggests that it is not a significant enough effect to fully resolve the discrepancy. 

The assumption of a fully-formed initial halo will also exacerbate this effect. More realistic mass aggregation histories are investigated by \scite{Firmani00} to the conclusion that, in earlier formed halos, gas cools at higher redshift when it is denser and consequently reaches equilibrium at smaller radii. In the light of this, future discussions on dynamics will be restricted to the correct hierarchical case (\S\ref{HierarchicalGrowth} \& \S\ref{Mergers}), leaving this isolated situation as merely a stage for introducing aspects of the disk's internal evolution.

\subsection{Disk Evolution}\label{DiskEvolution}
Once the disk has begun to form, the surface densities of gas, $\Sigma_{\rm g}$ and stars, $\Sigma_\star$ in a given annular region are evolved using a set of coupled differential equations. These describe the star formation and feedback process whereby supernovae both expel gas from the surrounding region and heat any which remains. The former effect depletes the fuel supply from which stars are created while the latter can promote the efficiency with which they form \cite{McKee77}. This is therefore a complex loop, controlled ultimately by the velocity dispersion of the existing clouds within the disk, $\sigma_{\rm g}$, and the infall of new gas, $\dot{\Sigma}_{\rm in}$ (due to the cooling of halo gas described in \S\ref{DiskFormation}). The entire system of equations describing this system is detailed below. Where necessary, we give alternate forms for three different models of star formation and feedback. The models used are:\\

\noindent {\bf (a):} The same fraction of gas is converted into stars every epicyclic period, a relationship which has empirical support \cite{Kennicutt98};\\

\noindent {\bf (b):} Star formation is promoted by gravitational instability, as proposed by \scite{Wang94}; \\

\noindent {\bf (c):} A model in which the star formation rate is proportional to the rate of gas cloud collisions.\\

\begin{equation}
\textrm{\underline{Star Formation Rate}}\label{stars}
\end{equation}
\begin{displaymath}
\dot{\Sigma}_\star = \epsilon_\star\frac{\Sigma_g}{\tau_\star}-f_{\rm r}m_{\rm SN}\dot{n}_{\rm SN}\hspace{1.35cm}\tau_\star=\left\{ \begin{array}{ll} 
\tau_{\rm o}  \hspace{0.5cm}& \textrm{(a)}\\ 
\\
\tau_{\rm e}\frac{Q}{\sqrt{1-Q^2}}  \hspace{0.5cm}& \textrm{(b)}  \\ 
\\
\frac{\left<\rho_{\rm cl}{r_{\rm cl}}^2\right>}{\Sigma_{\rm g}\sigma_{\rm g}} \hspace{0.5cm}& \textrm{(c)} 
\end{array}\right.
\end{displaymath} 
The parameter $\epsilon_\star=0.02$ is the fraction of gas converted into stars per orbital period, $\tau_{\rm o}$, as has been empirically determined \cite{Kennicutt98}. Some fraction of this initial mass will eventually be returned to the ISM through supernova explosions and stellar winds. A Kennicutt IMF gives this fraction to be $f_{\rm r}=0.41$  \cite{Cole00}, where the mass of stars formed per supernova $m_{\rm SN}=125{\rm M}_\odot$.

In model (b), the star formation rate is assumed to depend upon the {\it epicyclic} period,
\begin{equation}
\tau_{\rm e} = \left[\left.R^3\right/{\dif j^2\over\dif R}\right]^{1/2}
\end{equation}
and promoted by gravitational instability, as measured by the parameter
\begin{equation}\label{SFR(Q)}
Q =  \left[\pi G \tau_{\rm e}\left(\frac{\Sigma_g}{\sigma_g} + \frac{\Sigma_\star}{\sigma_\star}\right)\right]^{-1}.
\end{equation}
This implies that there is no star formation in stable disks ($Q>1$). The stellar velocity dispersion $\sigma_\star=\sigma_{\rm g}/5$.  

In model (c), the timescale for star formation, $\tau_\star$, is equal to the approximate time between collisions of molecular gas clouds, a possibility explored by \scite{Komugi06}. Basic kinetic theory for clouds of mass $m_{\rm cl}$ and radius $r_{\rm cl}$ suggests  
\begin{equation}
\tau_\star~\sim~\frac{\lambda_{\rm mfp}}{\sigma_{\rm g}}~\sim~\frac{m_{\rm cl}}{r_{\rm cl}\Sigma_{\rm g}\sigma_{\rm g}}\label{SFc} .
\end{equation}
The gas surface density is $\Sigma_{\rm g}$ and $\lambda_{\rm mfp}$ is the mean free path of the gas clouds. The size and density of molecular gas clouds were found by \scite{Dame86} to be related by  $\rho_{\rm cl} \approx r_{\rm cl}^{-1.3}$  which implies that the overall dependence on cloud properties in equation (\ref{SFc}) will be relatively weak: $\dot{\Sigma}_\star \propto r_{\rm cl}^{-0.7}$. With this in mind, and assuming that cloud sizes are fairly independent of the host system, eqn. (\ref{SFc}) leads to eqn.~(\ref{stars}c) by using characteristic values. The precise choice is not that important as a certain amount of uncertainty is implicitly contained in the free parameter $\epsilon_\star$. This work uses the value 
\[
\left<\rho_{\rm cl}r_{\rm cl}^2\right> = 2\times10^4M_\odot{\rm pc}^{-1}\hspace{.5cm}{\rm or}\hspace{.5cm}\tau_\star\approx \frac{2\times10^{10} {\rm years}}{\frac{\Sigma_{\rm g}}{M_\odot{\rm pc}^{-2}}\frac{\sigma_g}{{\rm kms}^{-1}}} . 
\] 
In this theory, star formation would be promoted by a high velocity dispersion. Equation~(\ref{SFR(Q)}) can be rearranged to show that, in (b), it would be promoted by {\it low} velocity dispersion so the two models are therefore in direct contrast in this respect. The comparative effect this has on the evolution of the system is discussed in \S\ref{IsolatedHalo}.\\

\begin{equation}
\textrm{\underline{Supernova Rate}} \label{SN} 
\end{equation}
\begin{displaymath}
\dot{n}_{\rm SN} = \int_0^{\tau_{\rm max}} \zeta(\tau)\dot{\Sigma}_\star(t-\tau) \dif\tau
\end{displaymath}
This calculates the current supernova rate per unit area, $\dot{n}_{\rm SN}$, from the star formation history. The function $\zeta(\tau)$ is the initial mass function expressed in terms of the stellar lifetimes (rather then, as usual, their mass). The limit on the integral, $\tau_{\rm max}\approx50{\rm Myr}$, is therefore the lifetime of the lowest mass star which will create a core-collapse supernova ($\approx8M_\odot$).\\

\begin{equation}
{\rm \underline{Conservation~of~mass}}\label{gas}
\end{equation}
\begin{displaymath} 
\dot{\Sigma}_{\rm g} = \dot{\Sigma}_{\rm in} - f_{\rm w}m_{\rm SN}\dot{n}_{\rm SN}   - \dot{\Sigma}_\star\hspace{0.6cm};\hspace{0.5cm}f_{\rm w} = \left\{ \begin{array}{ll} 
0.4  & \textrm{(a,b)}\\
\\
v_{\rm SN}/v_{\rm esc}& \textrm{(c)} 
\end{array}\right.
\end{displaymath} 
This equation incorporates four processes, the first of these being the infall of gas onto the disk, $\dot{\Sigma}_{\rm in}$ which is calculated from the technique described in section \ref{DiskFormation}. 

Gas is converted to stars, hence the term $-\dot{\Sigma}_\star$, and there will also be mass lost from the system due to winds, a process parameterised by $f_{\rm w}$. \scite{Marios06} examine a range of constant values for this parameter between 0 and 1 but a single value $f_{\rm w}=0.4$ is used here in the model which reflects their work (b), to avoid varying too many factors. This simple approach is also applied in model (a).

Model (c) introduces a functional dependance for $f_{\rm w}$ which is motivated by the following simple argument. If momentum is conserved perpendicular to the galactic plane in the case of a supernova releasing energy $E_{\rm SN}=10^{44}{\rm J}$ and ejecting mass $\Delta m_{\rm SN}=10{\rm M}_\odot$.
\begin{equation}
\dot{\Sigma}_{\rm out} < \frac{\sqrt{2E_{\rm SN}\Delta m_{\rm SN}}}{3v_{\rm esc}} \dot{n}_{\rm SN}<\frac{80{\rm kms}^{-1}}{v_{\rm esc}}m_{\rm SN}\dot{n}_{\rm SN}.
\end{equation}\label{wind}
Relativistic corrections to this argument would be small due to the low speeds involved and would only serve to lower the effective ejected mass, further enforcing the inequality. If this limit does indeed govern the process on an approximate level\footnote{Conserving energy for this process gives the bound:
\[
\dot{\Sigma}_{\rm out} < \frac{2E_{\rm SN}}{v_{\rm esc}^2}\dot{n}_{\rm SN} < \left(\frac{1000{\rm kms}^{-1}}{v_{\rm esc}}\right)^2 m_{\rm SN}\dot{n}_{\rm SN}.  
\]
So, even if as little as 3\% of this energy were actually available, the constraint of momentum conservation (\ref{wind}) is still the more relevant for $v_{\rm esc} < 1000{\rm kms}^{-1}$, which is the case for the majority of disks.}, equation (\ref{wind}) can be changed to the equality applied in equation (\ref{gas}c), hence $v_{\rm SN}=80{\rm kms}^{-1}$. \\
\begin{equation}
\textrm{\underline{Heating and cooling of the interstellar medium}}\label{sigma}
\end{equation}
\begin{displaymath}
\frac{\dot{\sigma}_{\rm g}}{\sigma_{\rm g}} = \epsilon_{\rm c}\frac{E_{\rm SN}\dot{n}_{\rm SN}}{\Sigma_{\rm g}\sigma_{\rm g}^2} -\frac{\Sigma_{\rm g}}{Q\tau_{\rm e}\Sigma_{\rm cool}} - \frac{\dot{\Sigma}_{\rm g}}{2\Sigma_{\rm g}}.
\end{displaymath}
This equation simply applies approximate conservation of energy to the disk of gas clouds.  Energy density, $\Sigma_{\rm g}\sigma_{\rm g}^2$, is gained from supernova explosions, reduced by the infall of cold halo gas and radiated away through collisions between gas clouds.

The expression for cooling rate follows \cite{Efstathiou00} and \scite{Marios06}, with all constants absorbed into the parameter $\Sigma_{\rm cool}=10M_\odot {\rm pc}^{-2}$. The value used by \scite{Marios06}  for the fraction of supernova energy absorbed, $\epsilon_{\rm c}=0.03$, is also followed here.

By making the working approximation $\dot{\Sigma}_\star\approx\left(~1-f_{\rm r}\right)m_{\rm SN}\dot{n}_{\rm SN}$, the interesting case where heating and cooling rates are equal can be evaluated analytically, giving an expression for the velocity dispersion at this point:
\begin{equation}\label{balance}
\sigma_{\rm eq}^2 = Q\epsilon_\star\epsilon_{\rm c}\frac{E_{\rm SN}}{m_{\rm SN}}\left(\frac{\tau_{\rm e}}{\tau_\star}\right)\frac{\Sigma_{\rm cool}}{\Sigma_{\rm g}}.
\end{equation}
This is plotted in Figs.~\ref{SingleHistories} \& \ref{SingleProfile} and used to help understand the results of the full dynamic calculation in terms of this simple approximation.
\pagebreak
\section{Applications}\label{Applications}

\begin{figure*}
\includegraphics[trim = 8mm 56mm 12mm 25mm, clip, height = 1.47\columnwidth]{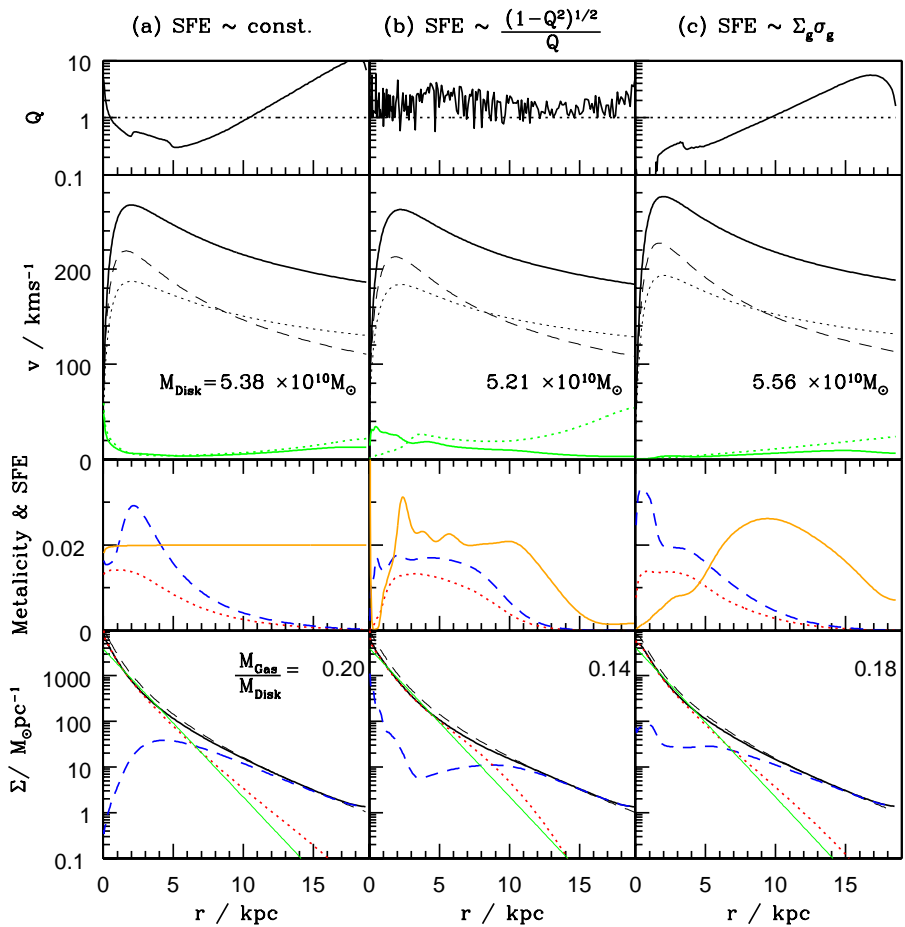}
\caption{{\bf Disks formed in isolation:  Radial profiles} of disks formed within an isolated halo with parameters as specified in the text. Columns correspond to the three different models of star formation described in \S\ref{DiskEvolution} (see labels at the top of each column). Throughout the Figures, dotted blue and dashed red lines represent the gaseous and stellar components respectively. Row~1 (the topmost) shows the gravitational stability parameter. In Row~2 the solid lines show the galaxy rotation curve. The contributions of the disk and the dark matter are indicated by faint dashed and dotted lines respectively. Green solid lines give the gas velocity dispersion, $\sigma_{\rm g}$ and dotted lines show the value that would be expected if heating and cooling rates were equal (see text). In Row~3 the star formation efficiency is shown by the solid yellow curve, while the others show the metallicities of gas and stars. Row~4 shows surface density profiles. The dashed black line indicates the total surface density in the absence of any outflow and the solid green lines show the exponential profile which best matches the stellar density. Scalelengths are (a) 1.34, (b) 1.34 and (c) 1.33 kpc. }
\label{SingleProfile}
\end{figure*}

\begin{figure}
\includegraphics[trim = 6mm 56mm 12mm 25mm, clip, width = \columnwidth]{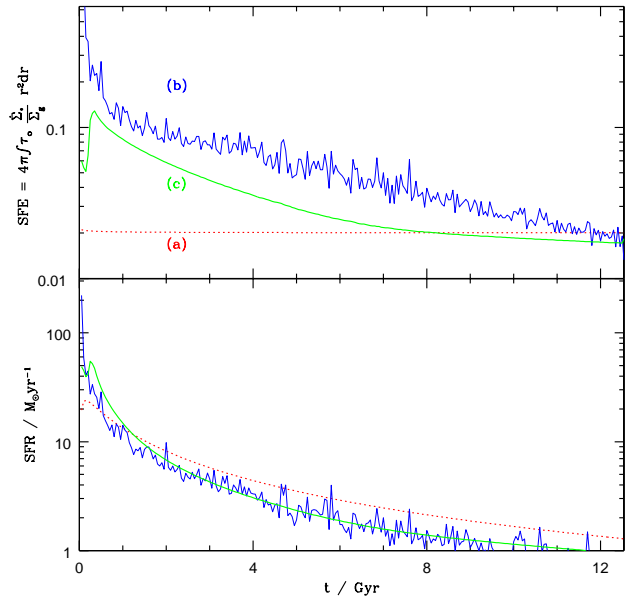}
\caption{{\bf Disks formed in isolation:  Star Formation Histories} of the systems from Figures \ref{SingleHistories} and \ref{SingleProfile}. The lower panel simply shows the total star formation in the disk as a function of time under each of the three models (a), (b) and (c). The upper panel shows a disk-averaged value for the effective star formation efficiency per orbit $\tau_{\rm o}\dot{\Sigma}_\star/\Sigma_{\rm g}$. }
\label{SFE}
\end{figure}

Firstly, in \S\ref{IsolatedHalo}, the model is used to examine whether it might be possible to identify the predominant mechanism of star formation directly from a disk galaxy's final state (Fig.\ref{SingleProfile}) or indirectly from its formation history (Figs.\& \ref{SFE} \ref{SingleHistories} ). For the purposes of this initial examination, and by way of introducing the model as a whole, disks are formed within an isolated halo which \emph{does not} undergo hierarchical growth but instead is assumed to have existed since $t=0$. The particular parameters of the halo are the same as were used to produce Figure \ref{alpha}.
Since one of the ultimate goals of this work is to assess the importance of hierarchical growth on disk formation models, sections \ref{HierarchicalGrowth} and \ref{Mergers} have been included to provide a brief illustration of its effect. Only one of the three models (a) is used for the purposes of studying this additional complexity, just as consideration of halo growth and mergers is set aside while investigating star formation mechanisms.

\subsection{An Isolated Halo}\label{IsolatedHalo}
\begin{figure*}
\includegraphics[trim = 9mm 56mm 12mm 25mm, clip, height = 1.47\columnwidth]{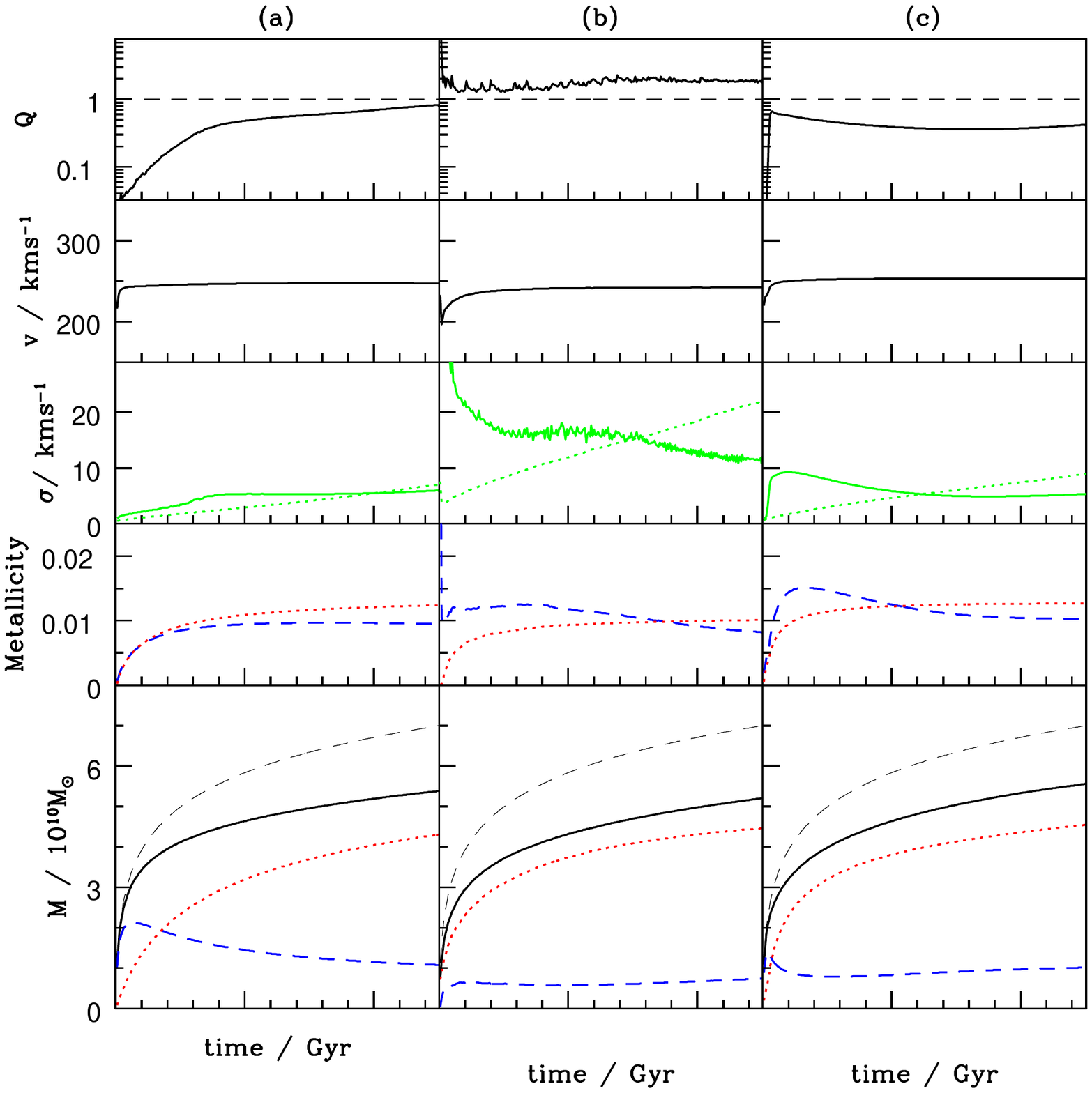}
\caption{{\bf Disks formed in isolation:  Histories} of the systems shown in Figure \ref{SingleProfile}. The cooling time at the virial radius is 12.5 Gyr. Row~1 shows mass-weighted mean gravitational stability and row 2 shows the circular velocity averaged over the stellar mass. The velocity dispersion averaged over the mass of gas is shown by the solid line in row 3 and the dotted line again represents the value that would be expected from energy balance. Row~4 shows the metallicities and row~5 shows the total masses, the dashed black line again representing the hypothetical case of no outflow. The bottom row shows the stellar disk scale length.}
\label{SingleHistories}
\end{figure*}
Many key features of the three resulting systems shown in Figure \ref{SingleProfile} show no clear evidence of the particular underlying physics. The considerably different assumptions made, notably regarding gas outflow, affect the overall mass distribution, and hence the circular velocity, by no more than 10\%. Consequently, the scalelengths of the exponential profiles which best match the stellar surface densities are also virtually identical, since nearly all the inner part of each disk has been converted into stars.

So, here are two principle observables of a disk galaxy, rotation curve and surface brightness profile, which can be explained equally well by three competing evolutionary theories. This is an ideal platform from which to ask what further observations would be required to break this degeneracy. The required details suggested by this simple study are the metallicity and velocity dispersion of the older (in this case inner) parts of the disk. 

The reason for differing velocity dispersions is most easily appreciated by considering energy balance in the ISM, an issue already alluded to in eqn.~(\ref{balance}). The velocity dispersion which corresponds to this state is shown by a dotted line in Fig.\ref{SingleProfile}, demonstrating that the value predicted by more dynamic modelling (solid line) is very close to this in the absence of recently infalling cold gas (which occurs mostly in the outer parts of the disk at this final stage).  The exception to this trend is the inner parts of the disk in model (b), where this equilibrium has evidently not been attained, recent star formation still heating the region effectively. 

In models (a) and (b) here, the star formation rate, and hence heating rate, is assumed to be proportional to epicyclic frequency. The cooling rate has a similar dependance so, under the condition of energy balance,
\begin{equation}
\sigma_{\rm g}\propto \Sigma_{\rm g}^{-\frac{1}{2}}.
\label{EnergyBalance}
\end{equation}
In the collisional model of star formation however, the heating rate is independent of dynamical timescale and the same condition leads to $\sigma_{\rm g}\propto\tau_{\rm e}$.  Since the epicyclic period is shorter toward the centre of the disk, this explains the lower central velocity dispersion in Figure \ref{SingleProfile}(c). This prediction is interestingly at odds with observations \cite{Kamphuis93,Schuster07} which find that $\sigma_{\rm g}$ monotonically decreases.

Taking a global view of the evolution, shown in Figure \ref{SingleHistories}, reveals that the systems cool below their equilibrium point after the initial burst of star formation and are kept at this lower level due to continuing infall of cold gas. This effect is most pronounced in (b).

All three theories predict similar eventual star formation efficiencies. However, the predicted histories differ considerably, as can be seen in Figure \ref{SFE}. The clearest feature is that a dynamic star formation mechanism could produce much higher initial efficiencies whilst still being consistent with the relative inefficiency which is observed at the present-day . 

According to model (b), this behavior is caused by high instability in the embryonic disk. As the cold gas is depleted, the disk stabilizes and the star formation efficiency duly drops to its final value. The evolution of the stability parameter $Q$, along with other quantities, can be seen in Figure \ref{SingleHistories}.  

Under the assumptions of model (c), it is the frequent cloud collisions in the highly dense, early formed inner regions which cause the high efficiency. Subsequent loss of efficiency as the gas density drops is exaggerated further by the accompanying drop in velocity dispersion (Fig.\ref{SingleHistories}) as supernova heating decreases along with star formation. This is why the third model tends more rapidly to a low efficiency.

All these differences remain fairly subtle, suggesting that the final state of a galactic disk, however uniquely it may have been reached, is predominantly governed by gas supply, gravitational dynamics and stellar lifetimes, which are the same for all the models presented here. The effects of hierarchical growth, as shown in the next section, would swamp most signatures of the star formation process.

Other models introduce bulge formation through disk instability. Van den Bosch (2001) examines the fractional contribution made by the disk to the circular velocity $v_{\rm disk}/v$ and assumes that disks where this exceeds 0.7 at any point are globally unstable. The disk component of these systems is indeed larger than this value, but only marginally (just 5-10\% of the mass would have to be redistributed in order to satisfy the criterion). \scite{AR00} assume instead that stars from regions exceeding a critically stable surface density, $\Sigma_\star=\sigma_\star/3.36\tau_{\rm e}{\rm G}$, are transfered from the disk to the bulge until the equality is satisfied. This would have a very dramatic effect on the system in this example, removing much of the inner part of the disk ($r<5{\rm kpc}$) and leaving a remainder with little resemblance to an exponential profile.

A clear set of theoretical or observational evidence is still needed to support one particular analytic approach to this issue. In lew of this, we note the theory that perturbations on scales larger than $\lambda_{\rm c} = 4\pi^2G\Sigma\tau_{\rm e}$ will not propagate even if $Q>1$ \cite{Binney87}. In this particular example,  $\lambda_{\rm c}$ is never more than about a tenth of that region's galactic radius and is usuallly much less.

\subsection{Hierarchical Growth}\label{HierarchicalGrowth}
\begin{figure}
\includegraphics[trim = 8mm 56mm 70mm 25mm, clip, height = 1.47\columnwidth]{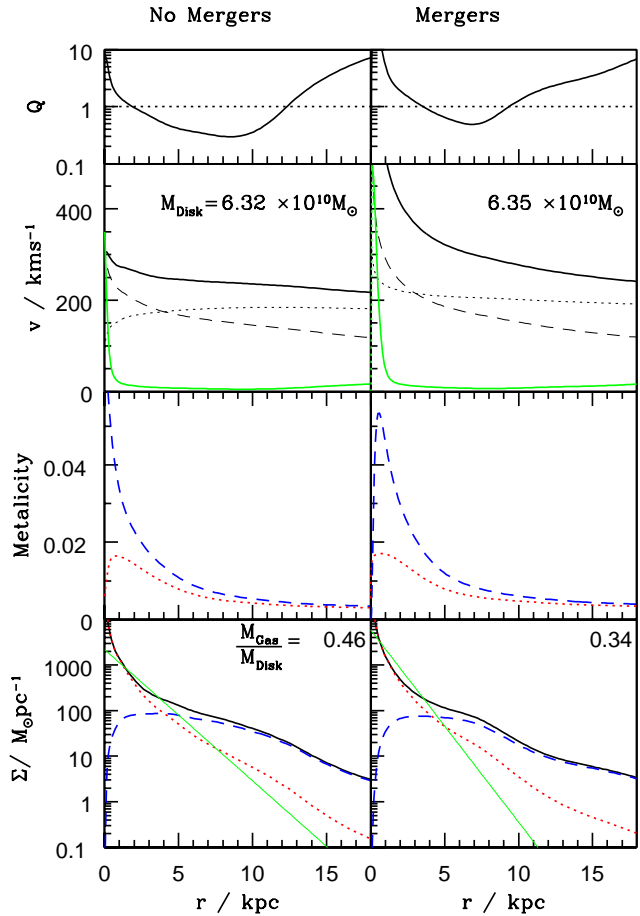}
\caption{{\bf Column 1-- Hierarchical formation with no galaxy merging:} See the caption to Fig.~\protect\ref{SingleProfile} for the key. The exponential scalelength which best fits the stellar surface density is 1.5kpc.   {\bf Column 2---Galaxy merging Included:} . Following the formation of a central bulge with mass $2.9\times10^{10}M_\odot$, the stellar disk scalelength reduces to 1.0kpc. The dashed black line in the plot velocity vs. radius still shows just the disk's contribution to the rotation curve.}\label{FullProfile}
\end{figure}
\begin{figure}
\includegraphics[trim = 8mm 56mm 70mm 25mm, clip, height = 1.47\columnwidth]{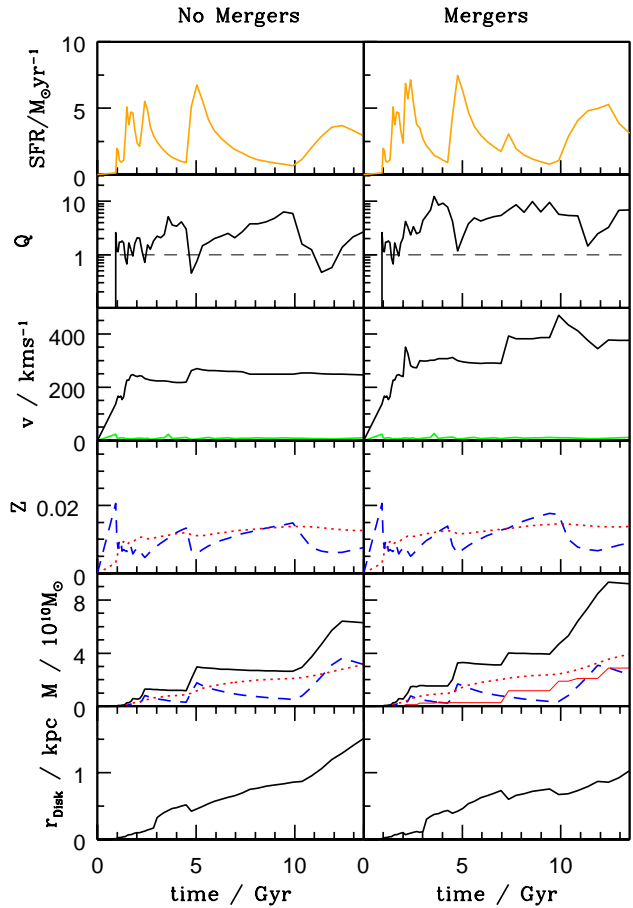}\label{History1}
\caption{{\bf Column 1-- Hierarchical formation with no galaxy merging:} See the caption to Fig.~\protect\ref{SingleHistories} for the key.   {\bf Column 2 -- Galaxy merging Included:}. The growing mass of the bulge component is shown here by the solid red line in the plot of Mass vs. time, culminating in a final Bulge/Disk ratio of 0.45. The dotted red line still shows the total stellar mass, which now includes the stars in the central bulge.}\label{FullHistory}
\end{figure}
The calculation of disk formation is now embedded within the {\sc Galform} semi-analytic framework. A statistically representative sample of dark matter halo merger trees was generated, corresponding to present-day halo masses in the range $10^{12}$ to $10^{13}h^{-1}M_\odot$. From these disk-dominated galaxy at $z=0$ has been selected which shares similar properties with the isolated system in \S\ref{IsolatedHalo}. These calculations used the standard {\sc Galform} assumptions, adopting parameters chosen by \scite{Baugh05}.

A full description of the hierarchical growth of structure, and how this affects the formation of galaxies, can be found in \scite{Cole00}. We will briefly discuss aspects of this implementation central to the present work. 

The growth of structure in the dark matter component is followed by constructing merger trees on a fine grid of timesteps. Within this merging hierarchy of halos we then proceed to follow galaxy formation. The first halos to form are assumed to have a fraction $\Omega_{\rm b}/\Omega_0$ of their mass in the form of baryons which are shock heated to the halo virial temperature. We then track the cooling radius within this atmosphere of hot baryons (see \S\ref{Halo}) to determine the rate at which gas cools. Two aspects of this cooling are crucial for the present work:\\

\noindent(i) Later generations of halos will contain a smaller fraction of baryons in the hot atmosphere as some will have been locked up into galaxies at earlier times;\\

\noindent(ii) The growth of halo mass is a continuous process, but within {\sc Galform}, we impose discrete halo formation events which occur whenever a halo has doubled in mass since the previous formation event. At such events, the temperature of the hot gas is reset to the new virial radius and the cooling radius is reset to zero (beginning to grow again as described in \S\ref{Halo}).\\ 

The second point has been a traditional feature of semi-analytic models (originally arising for pragmatic
reasons). It is retained in this work in order to maintain similarity to previously published work but its lack of physical motivation is acknowledged and a more realistic cooling model (such as that of \pcite{McCarthy07}) is planned.

In the first example here, illustrated in the first column of Figures \ref{FullProfile} and \ref{FullHistory}, the effects of mergers between galaxies are ignored. Specifically, the effects of dynamical friction are not included, so sub-halos and their galaxies will continue to orbit within their host halo forever. As such, no elliptical galaxies or bulges can form\footnote{Note that the \protect\scite{Baugh05} model does not allow for the formation of spheroidal systems through global instabilities of disks.}. 

Several differences with respect to the isolated case are immediately apparent. Firstly, star formation now occurs in short bursts, even though the star formation is occurring purely in disks. This is due to the hierarchical build-up of structure---significant merger events replenishing the supply of hot gas which is available to cool onto a forming disk. (This also causes a decline in the gas metallicity.) This correspondence can be seen directly by comparing the times of peak star formation rate with the step-like increases in disk mass and rotation speed. It should be noted that observations of the Milky Way are consistent with a discontinuous star formation history \cite{Pinto00}.

Though qualitatively similar to the rotation curves shown in \scite{Firmani00} and \scite{vdB02}, those predicted by this model are definitely less consistent with observations, the rotation speed being higher in the disk centre for reasons discussed in \S\ref{DiskFormation}.

After an initial high peak, the velocity dispersion of the hierarchically-formed systems tends to increase slightly with radius, as would be expected from the approximate dependance in equation (\ref{EnergyBalance}). This is again in marginal disagreement with observations, suggesting that this evolution model may be too simplistic.

The surface density profiles of the stellar disks are no longer completely smooth, due to the discontinuous nature of hierarchical growth. This problem is believed to be exacerbated somewhat by the nature of the cooling model used in {\sc Galform}. As described by \scite{Cole00}, the cooling time for hot halo gas in {\sc Galform} is reset to zero each time a halo is considered to have ``formed'' (formation events are taken to occur whenever the halo's mass has doubled since the previous formation event).

\subsection{Galaxy Mergers}\label{Mergers}

Finally, we allow galaxy-galaxy mergers to occur and, in the case of major mergers, to result in the formation of spheroidal systems which may later become the bulges of disk-dominated galaxies. In the {\sc Galform} model, satellite galaxies within a halo can lose orbital angular momentum via dynamical friction and eventually merge with the galaxy residing at the centre of that halo. The results of such mergers are found by applying the usual rules of {\sc Galform} (see \scite{Cole00} for details).

If disks are destroyed by mergers, we use the properties at the half-mass radius of the disk (now determined explicitly by our calculations) in determining the properties of the resulting elliptical galaxy. In minor mergers, disks are not destroyed but added together by assuming that the cumulative mass as a function of specific angular momentum is the same for both the combined two-disk system and the resulting, single disk. The new radial density profile is then determined as described in \S\ref{DiskFormation}. 

In the presence of a bulge, eqn.~\ref{R_basic} is modified to become
\begin{equation}
R = {j^2 \over {\rm G} [M_{\rm disk}(j)+M_{\rm DM}(j)+M_{\rm bulge}(R)]}
\end{equation}
and is solved iteratively. $M_{\rm bulge}(R)$ is the mass of the bulge within radius $R$. We assume a de Vaucouler's profiles for bulges, and solve for the characteristic radius using the same methods as \scite{Cole00}.

This treatment neglects the orbital angular momentum possessed by the galaxies just prior to merging. We are not aware of any detailed (e.g. N-body) experiments which have assessed how this angular momentum is redistributed into the resulting galaxy (or if, instead, it is mostly lost from the system), although a merger-driven disk formation scenario has been proposed by \scite{Robertson06}. 

To quantify the importance of this orbital angular momentum, we performed a simple test. Specifically, prior to merging the satellite galaxy onto the primary galaxy we incresed the specific angular momentum of each annulus of material in the satellite disk by an amount equal to its orbital specific angular momentum. We found essentially no change in the resulting profiles of disks at $z=0$ and therefore conclude that a correct treatment of this orbital angular momentum would likewise cause only minor changes in the predicted galaxy profiles.

The resulting galaxy disks shown in the second column of Figures \ref{FullProfile} and \ref{FullHistory} are very similar to the mergerless case, despite the fact that approximately 30\% of the final galaxy mass is in a bulge component. One significant and expected difference is in the rotation curve which rises even more rapidly towards the centre due to the additional central mass.

\section{Conclusions}\label{Conclusions}

We have described methods for expanding the range of physical properties predicted by semi-analytic models of galaxy formation to include the surface density profiles of stars and gas in galaxy disks. These profiles are computed self-consistently from the assumed distribution of angular momentum in the hot halo gas which fuels galaxy formation. 

Using the additional information provided by these resolved disks we are able to explore more detailed models of star formation than has previously been possible within semi-analytic models.  One key difference is the provision for dynamic evolution of the system, free from any prior enforcement of fixed stability \cite{Firmani00,Efstathiou00} or energy balance \cite{vdB01,Marios06}.

These methods have been implemented within the {\sc Galform} model in order to study how disks form and grow in a hierarchical Universe, exploring three cases of increasing complexity and realism.

The main conclusion from this initial, brief study is that the primary factor governing the evolution of these systems is the availability of gas for star formation, the introduction of hierarchical growth dramatically altering the form of the resulting star formation histories. This emphasizes the fact that any realistic model of disk formation \emph{must} be embedded within the appropriate hierarchically growing distribution of dark matter halos.

By comparison, the structure and internal properties of the final disk systems are relatively insensitive to the assumptions concerning star formation. The chemical evolution and star formation histories exhibit greater distinction, so this conclusion can not necessarily be extended to galaxies of all masses at all redshifts---further work is needed to explore these regimes. 

There is no doubt that a prescription for star formation could be created which \emph{did} produce contrasting results, but the fact that three different, physically motivated star formation models produce quite similar present-day systems suggests that semi-analytic models (and hydrodynamical simulations which frequently use similar rules for the sub-grid star formation physics) can produce robust predictions for galaxy properties even though the precise rules governing star formation are not known. 

Another key result is that current estimates of the distribution of angular momentum in halo gas from up-to-date numerical simulations \cite{Sharma05} correspond to an excessive quantity of low angular momentum material, consistently producing disks whose rotation curves rise too sharply in the centre, a problem which has been well noted \cite{Bullock01,vdB01b,vdB02b,Maller02}. Even more accurate characterisations of the distribution of halo gas angular momentum, and further studies of how this is conserved (or otherwise) during cooling and star formation are keenly awaited.

While this work explores the evolution of a single system, it is just one of a large, statistically representative sample of galaxies forming in halos of a wide range of masses. The full potential of this modeling technique will be exploited in forthcoming work, its predictions being set against both cosmological data-sets and collected galactic profiles (of stellar and gas mass, rotation speed, age and metallicity). This will provide an important assessment of the compatibility between $\Lambda$CDM cosmology, as represented by its predictions for large scale structure formation, our understanding of star formation and the observed Universe.
\newpage
\section*{Acknowledgments}

We would like to thank Carlton Baugh, Richard Bower, Shaun Cole, Carlos Frenk, John Helly, Cedric Lacey and Rowena Malbon for allowing us to use the {\sc Galform} semi-analytic model of galaxy formation ({\tt www.galform.org}) in this work. We would also like to thank Joe Silk, Marios Kampakoglou and Mark Kamionkowski for valuable discussions and the anonymous referee for detailed comments which led to considerable improvements in the manuscript. 

AJB acknowledges support from the Gordon and Betty Moore Foundation. MJS acknowledges a PPARC studentship at the University of Oxford and the hospitality of the Center for Theoretical Cosmology and Physics at Caltech where much of this work was completed.

\appendix

\section{Numerical Robustness}\label{CompDetails}
As described in \S\ref{DiskFormation}, processes internal to each disk are followed for each of 200 concentric annuli and external processes, e.g. galaxy mergers, are followed on a coarser set of 20 ``zones" (also concentric annular regions, but a different term is used to avoid confusion).

\begin{figure}
\centering
\subfigure{\includegraphics[trim = 7mm 40mm 12mm 10mm, clip, height = 1.2\columnwidth]{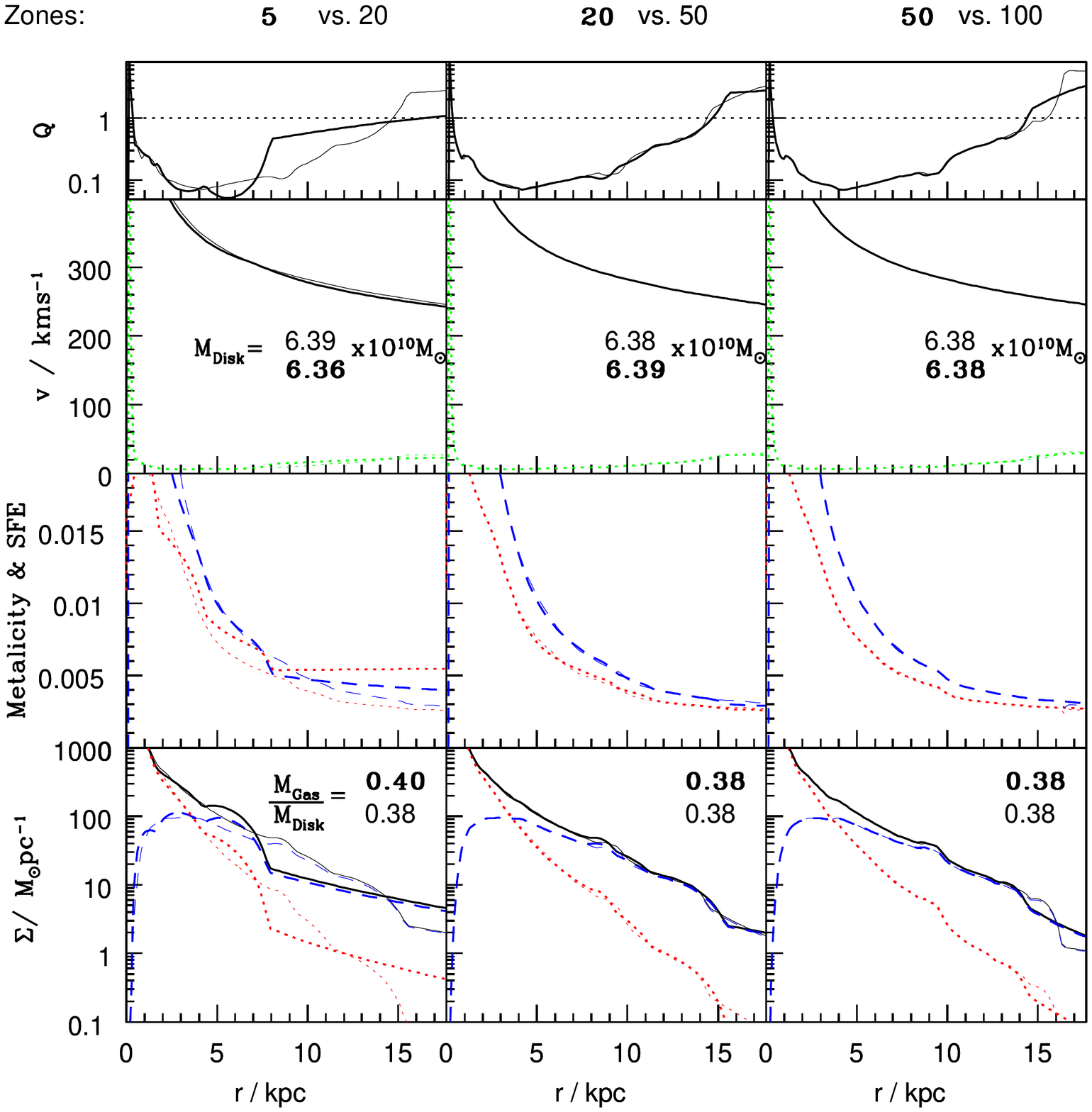}}
\subfigure{\includegraphics[trim = 7mm 40mm 12mm 10mm, clip, height = 1.2\columnwidth]{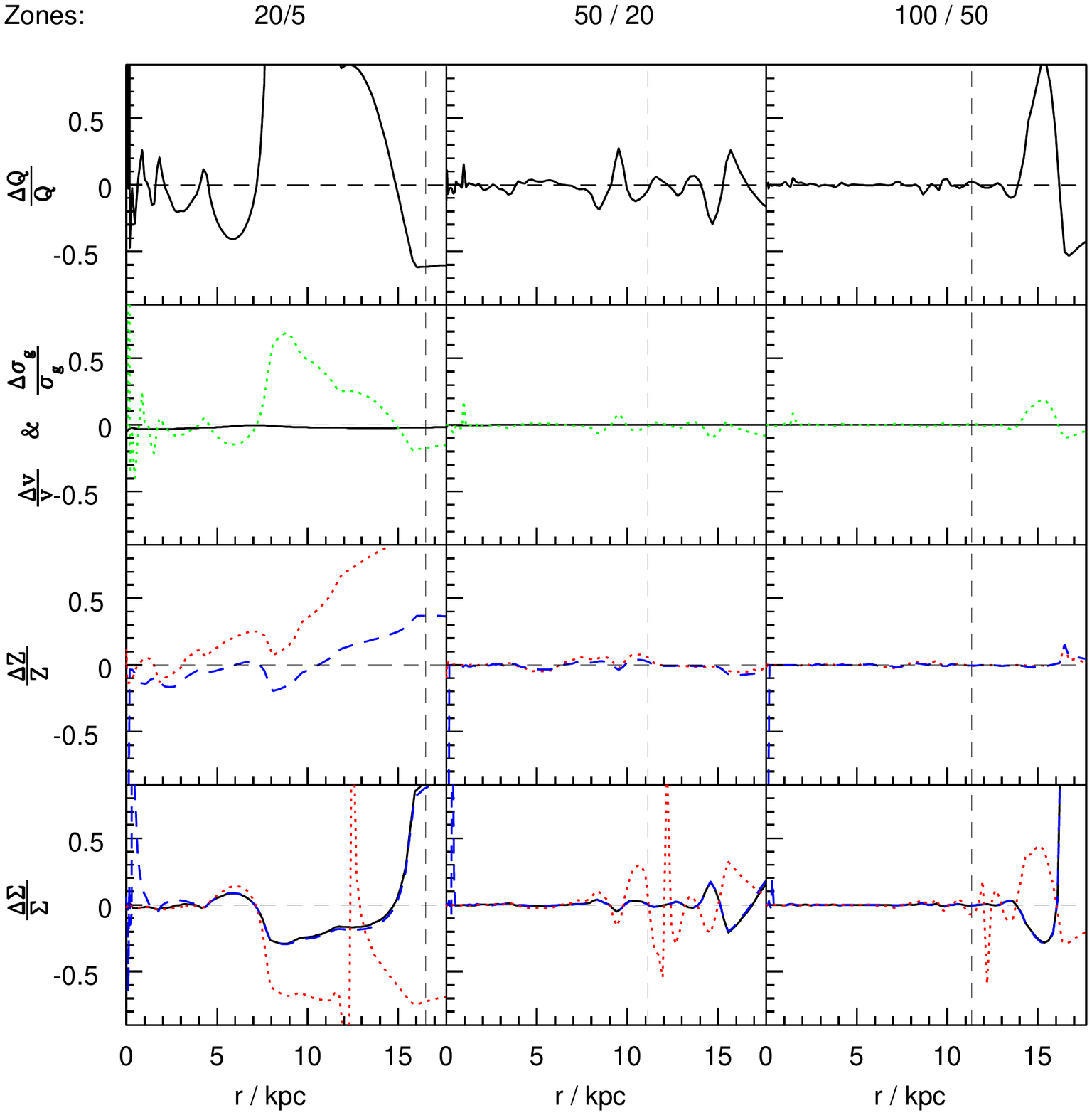}}
\caption{\emph{Left panel:} Radial profiles of the galaxy examined in \S\ref{Applications} under model (a), as generated using different numbers of radial divisions used in calculations of mergers (``zones''). Lines have the same meaning as in Fig.~\protect\ref{SingleProfile}. The disk mass and gas fractions given in the panels correspond to the number of Zones with the corresponding typeface.The vertical dotted line encloses 90\% of the disk's mass. \emph{Right panel:} Fractional differences between each pair of lines.}
\label{Zones}
\end{figure}

\begin{figure}
\centering
\subfigure{\includegraphics[trim = 7mm 40mm 70mm 10mm, clip, height = 1.2\columnwidth]{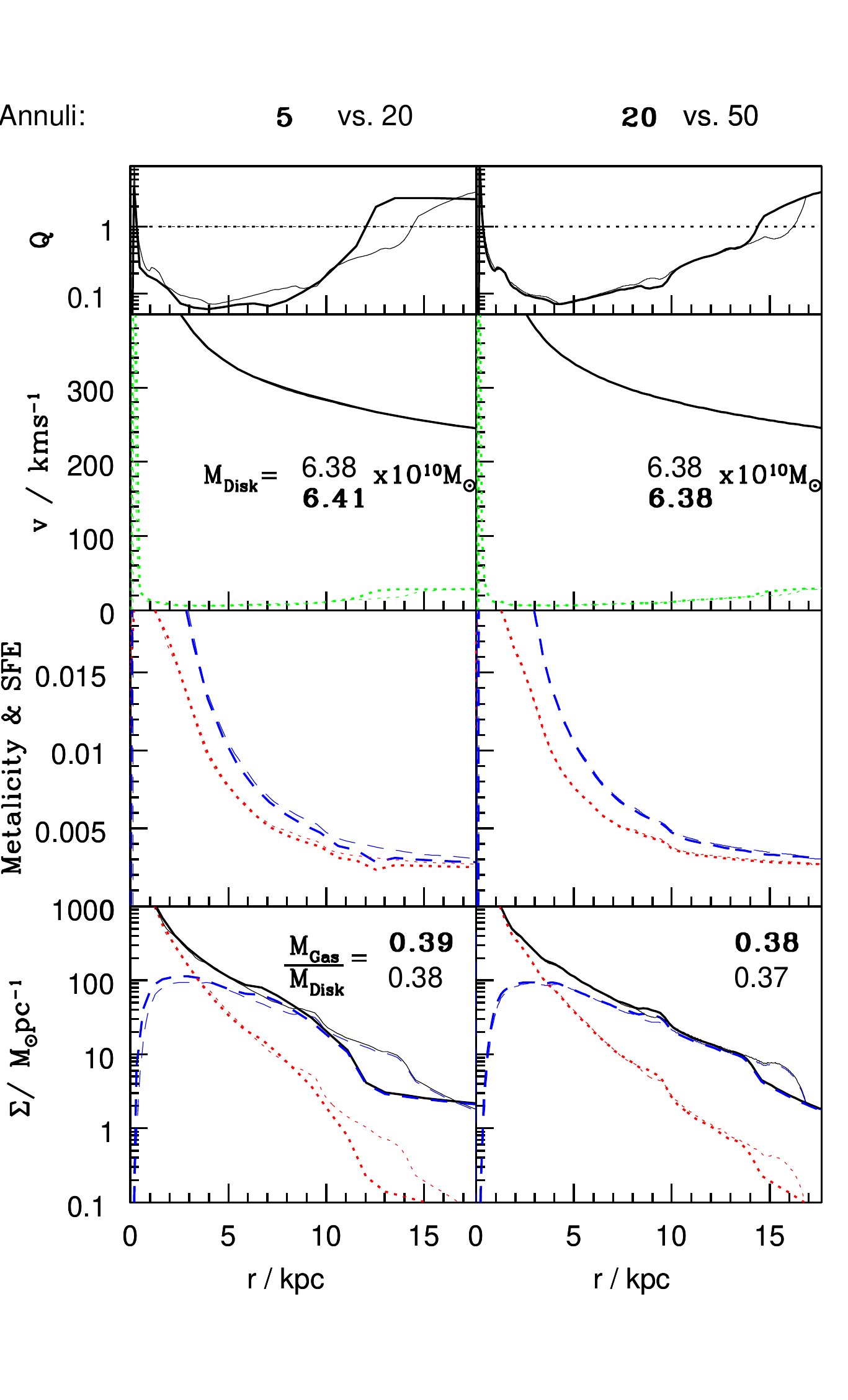}}
\subfigure{\includegraphics[trim = 7mm 40mm 70mm 10mm, clip, height = 1.2\columnwidth]{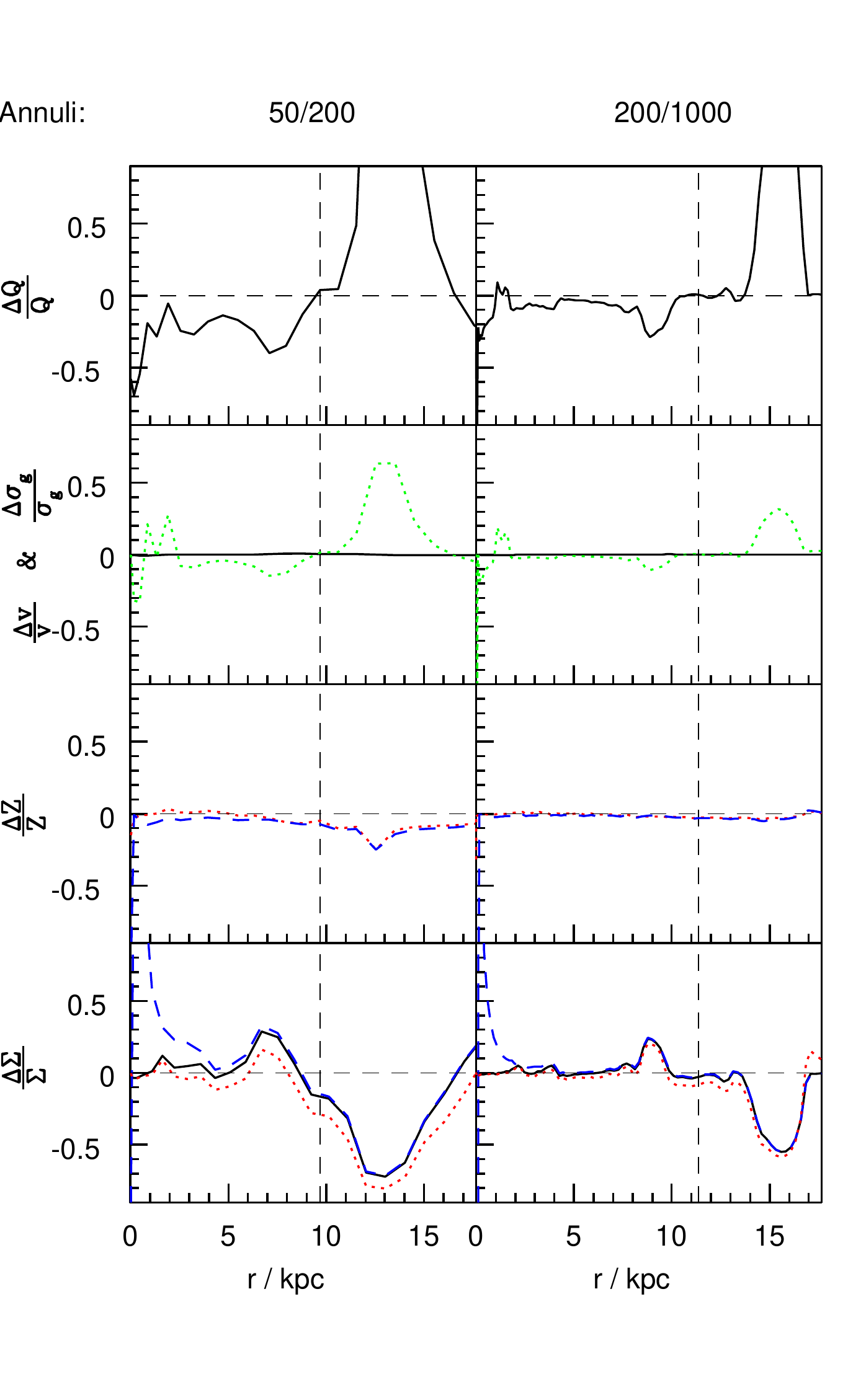}}
\caption{As for Figure \ref{Zones}, but now varying the number of radial divisions used in calculations concerning the internal evolution of each galaxy (``annuli''). The vertical dotted line again encloses 90\% of the disk's mass.}
\label{Residuals}
\end{figure}

To establish the numerical robustness of these approximations, we have repeated our calculations using different numbers of annulli and zones. 
Figure~\ref{Zones} shows the effect these alterations had on the eventual profile of the typical galaxy examined in this paper and Figure \ref{Residuals} shows the fractional changes in each parameter.

It is representative of the finding, for all of the modelled galaxies, that increasing the number of zones above 20 made no more than a 5\% difference in any eventual parameter value within the inner 90\% of the disk's mass. In general, the difference was much less than this. Our calculations are therefore relatively unaffected by these numerical choices.

\end{document}